\documentclass[twocolumn, nofootinbib,nobibnotes, showpacs]{revtex4}
\usepackage{amsmath,amssymb,amsfonts,times,graphicx,color}
\usepackage{esint}

\usepackage{hyperref}
\usepackage{eucal}
\definecolor{darkred}{rgb}{0.8,0.1,0.1}
\hypersetup{colorlinks=true, linkcolor=darkred, citecolor=blue}

\begin{document}

\title{Entanglement Entropy from the Truncated Conformal Space}

\author{T. Palmai}
\email{palmai@phy.bme.hu}

\affiliation{\textit{\small MTA-BME "Lendulet" Statistical Field Theory Research Group, Institute of Physics, Budapest University of Technology and Economics, Budafoki ut 8, 1111 Budapest, Hungary}}
\begin{abstract}
A new numerical approach to entanglement entropies of the R\'enyi type is proposed for one-dimensional quantum field theories. The method extends the truncated conformal spectrum approach and we will demonstrate that it is especially suited to study the crossover from massless to massive behavior when the subsystem size is comparable to the correlation length. We apply it to different deformations of massless free fermions, corresponding to the scaling limit of the Ising model in transverse and longitudinal fields. For massive free fermions the exactly known crossover function is reproduced already in very small system sizes. The new method treats ground states and excited states on the same footing, and the applicability for excited states is illustrated by reproducing R\'enyi entropies of low-lying states in the transverse field Ising model.
\end{abstract}


\maketitle

\section{Introduction}

Entanglement is a fundamental quantity characterizing correlations in quantum field theories and quantum many-body systems. Quantifying entanglement in ground states can be useful to detect and describe phase transitions, even when a conventional order parameter is unavailable \cite{Amico2008}, while excited state entanglement can shed light on spreading of correlations in out of equilibrium time evolution (see e.g. \cite{quench}).

Let us consider a one-dimensional system and examine entanglement entropies relative to a bipartition as a function of the subsystem size. When the subsystem size is much smaller than the correlation length, the model becomes effectively gapless. Entanglement entropies in this case provide universal information about the underlying conformal field theory: in the ground state a logarithmic law proportional to the (effective) central charge is observed \cite{HLW1994,CC2004,Bianchini2015}. In case of excited states, the leading behavior for small subsystems is unchanged, but corrections appear, in scrutiny with the operator content of the theory, e.g. scaling dimensions of the most relevant operators can be identified \cite{ABS2011}. In fact, since conformal field theories are exactly solvable, entanglement entropies are calculable for any state in the conformal space and recent studies have been concentrating on this aspect to understand properties of excited states in critical systems \cite{P2014,CVO2015,CGHW2015}.

If the subsystem size is much larger than the correlation length, the area law governs entanglement in the ground state of local Hamiltonians: entanglement entropy of a bipartition scales with the size of the boundary. In one spatial dimension this predicts a constant in the subsystem size \cite{arealaw}.

In this paper we will be interested in the crossover between these two regimes. It was already shown that the crossover function is very interesting and encodes certain, generic information about the system, e.g. the first correction to the area law is determined only by the masses of stable particle excitations in the massive model \cite{CCAD2008,D2009}. Our aim here will be to develop and test a new approach capable to capture the whole scaling function describing the crossover from gapless to gapped behavior both for the ground state and excited states. We will work in relativistic quantum field theories, the scaling limit of many interesting 1D systems. Our approach combines two established methods: first we will use the \emph{truncated conformal space} to expand a state in the massive theory on the UV limiting conformal field theoretical space of states, and then we calculate the R\'enyi entanglement entropy exploiting conformal symmetry.

Our approach essentially extends the truncated conformal space approach (TCSA) to R\'enyi entropies. The TCSA \cite{tcsa} is a Hamiltonian truncation approach that enables to find states in the UV limiting CFT Hilbert space that correspond to ground states (or on the same footing, excited states) of massive deformations of the CFT through the quantum mechanical variational principle. A finite matrix problem is obtained by going into finite volume and imposing an energy-wise truncation on the basis states. The truncation is justified by the expectation that for relevant perturbations and in finite volume the high energy sectors of the perturbed and unperturbed theories are the same and they do not couple to the low-energy sector. Diagonalizing this finite Hamiltonian matrix numerically, yields the spectrum of the system but also gives the corresponding eigenstates in terms of CFT states. The TCSA has proven to be a powerful and versatile tool to evaluate not only the spectrum \cite{tcsa} but various other quantities depending on the expansion of the states, e.g. form factors \cite{PT2008} and boundary g-functions \cite{TW2012}. We are encouraged by this success and indeed we will find that also in the case of R\'enyi entropy the TCSA is very robust.

Via TCSA we reduce the problem of finding the R\'enyi entropy of an arbitrary QFT state to  finding it for an arbitrary CFT state. The advantage of this approach is that it enables to work in CFT and exploit conformal symmetries in the calculations. 
For CFT states R\'enyi entropies can be calculated efficiently by the replica approach. In this we use the fact that the $n$th R\'enyi entropy of an arbitrary state can be represented as a $2n$-point functions on a Riemann surface, which in CFT can be mapped to $2n$-point functions on the plane. These objects can be evaluated using standard methods \cite{DiFrancesco}. The replica method was initially described for the ground state of a CFT \cite{HLW1994,CC2004} and later it was generalized to states corresponding to primary fields \cite{ABS2011} and more recently also to arbitrary elements of the Hilbert space \cite{P2014}. Our strategy is to use this latter generalization in conjunction with the TCSA.

The advantage of our approach is at least two-fold.
First, TCSA can treat a diverse array of models, including both integrable and nonintegrable ones, scaling limits of spin chains, through bosonization models relevant in strongly correlated electron physics \cite{bosonize} or treating the interchain coupling as a perturbation can even be useful to understand two-dimensional systems \cite{2Dtcsa}. And second, finite volume effects are exponentially suppressed in field theory \cite{Luscher}, making our approach a useful alternative to lattice methods when studying the thermodynamic limit.

To demonstrate these features we study two perturbations of the massless Ising model, implemented by the energy density operator giving rise to the massive Ising model, and by the magnetization operator corresponding to Zamolodchikov's famous $E_8$ model \cite{Zamolodchikov1991}. We will see that already in very small volumes the infinite volume behavior is recovered with great precision and the results quickly become volume independent.

The paper is organized as follows. In Section 2 previous theoretical results are recalled about R\'enyi entropies in quantum field theories. In Section 3 we outline the idea of the TCSA and its application to calculate R\'enyi entropies. In Section 4 we present results on deformation of the massless $c=1/2$ Ising field theory, reproduce the second R\'enyi entropies of various low-lying states in the massive Ising field theory and its lattice realization, the TFIM, and study the effects of the longitudinal field. Section 5 is reserved for conclusions.

\section{R\'enyi entropy in quantum field theory}

In this paper we will concentrate on the second R\'enyi entropy, a particular measure of entanglement relative to a spatial bipartition of the system, $A\cup B$. It and can be defined as
\begin{equation}
S_A^{(2)}=-\log\text{Tr}_A \rho_A^2,\qquad \rho_A=\text{Tr}_B\rho
\end{equation}
with $\rho=\vert\Psi\rangle\langle\Psi\vert$ full and $\rho_A$ partial density matrices. $\vert\Psi\rangle$ is a particular state of the full system. $S_A^{(2)}$ is also equivalent to the purity of the subsystem $A$.

In a field theoretical setting it is useful to reinterpret the quantity $\text{Tr}_A\rho_A^2$ in terms of a Euclidean path integral \cite{HLW1994,CC2004}. We find that the field in the path integral lives on a Riemann surface $\mathcal{R}$ with nonzero curvature, in case of the ground state,
\begin{equation}\label{PIRiemann}
\text{Tr}_A\rho_A^2\propto \int [d\varphi]_\mathcal{R}\exp\left[-\int_\mathcal{R}d\tau dx \,\mathcal{L}[\varphi](\tau,x)\right]
\end{equation}
where the proportionality reflects a further normalizing constant setting $\text{Tr}\rho=1$. The Riemann surface $\mathcal{R}$ is composed of two sheets sewn together along the subsystem $A=(0,a)$ at Euclidean time $\tau=0$ according to 
\begin{eqnarray}
(-0,A_1)\leftrightarrow (+0,A_2)\\
(-0,A_2)\leftrightarrow (+0,A_1)
\end{eqnarray}

We can obtain a local theory by rewriting this path integral in terms of two copies of the original field both living on the Euclidean plane $\mathbb{R}^2$, however at a price of inserting two local operators into the integral,
\begin{multline}\label{PITwist}
\text{Tr}_A\rho_A^2\propto \int [d\varphi_1d\varphi_2]_{\mathbb{R}^2}\mathcal{T}(0,0)\tilde{\mathcal{T}}(0,a)\times\\
\times\exp\left[-\int_{\mathbb{R}^2}d\tau dx \,(\mathcal{L}[\varphi_1]+\mathcal{L}[\varphi_2])(\tau,x)\right]
\end{multline}
and the local operators, dubbed as branch-point twist fields (for the second R\'enyi entropy $\mathcal{T}(\tau,a)=\tilde{\mathcal{T}}(\tau,a)$), implement the following conditions
\begin{align}
\mathcal{T}(\tau,a):\quad&\varphi_1(\tau+0,x)=\varphi_2(\tau-0,x)\\
&\varphi_2(\tau+0,x)=\varphi_1(\tau-0,x),\quad x\in[a,\infty)
\end{align}
In this way, we obtained the exponentiated R\'enyi entropy as a two-point function of the twist fields,
\begin{equation}
\text{Tr}_A \rho_A^2=N\langle\mathcal{T}(0,0)\tilde{\mathcal{T}}(0,a)\rangle_{\mathcal{L}[\varphi_1]+\mathcal{L}[\varphi_2],\mathbb{R}^2}
\end{equation}

The formalism generalizes to excited states by replacing in the previous equation the vacua by the excited states,
\begin{equation}
\text{Tr}_A \rho_{\Psi,A}^2=N\langle\Psi\vert\mathcal{T}(0,0)\tilde{\mathcal{T}}(0,a)\vert\Psi\rangle_{\mathcal{L}[\varphi_1]+\mathcal{L}[\varphi_2],\mathbb{R}^2}
\end{equation}

\subsection{Massive theories}

To calculate the two point function of twist fields in massive quantum field theories a form factor approach was developed, initially in the presence of integrability \cite{CCAD2008}.

The most important result of this program was to determine the entanglement entropy's ($S_A=-\text{Tr}_A(\rho_A\log\rho_A)$) approach of the area law when the correlation length and the subsystems size is of the same order of magnitude. The formula is quite generic and applies irrespective of integrability \cite{D2009},
\begin{equation}\label{SA}
S_A=\frac{c}{3}\log{\xi}+U-\frac{1}{8}\sum_{\alpha=1}^\ell K_0(2rm_\alpha)+O(e^{-3rm})
\end{equation}
where $\xi\sim1/m_1$ is the correlation length, $U$ a nonuniversal constant, $K_0$ is the modified Bessel function. In the slowest decaying term (in terms of the subsystem length $r$), interestingly, the spectrum of stable particle excitations $m_\alpha$ enter \cite{CCAD2008,D2009}. The nonuniversal constant is such that for $r\to0$ one would recover the CFT result.

For the R\'enyi entropy the leading behavior is the same, but with different constants,
\begin{equation}\label{S2A}
S_A^{(2)}=\frac{c}{4}\log{\xi}+U^{(2)}-\kappa^{(2)}\sum_{\alpha=1}^\ell K_0(2rm_\alpha)+O(e^{-3rm})
\end{equation}
where $U^{(2)}=-2\log\langle\mathcal{T}\rangle$ comes from the vacuum expectation value of the twist field.  A less suppressed decay of $e^{-rm}$ is also believed to be possible for the R\'enyi entropies, when one-particle form factors of the twist field are non-vanishing \cite{CCAD2008}, however such decay was not yet observed.

When the theory is integrable in principle the form factors can be determined exactly by solving the form factor bootstrap and further corrections to (\ref{SA}) and (\ref{S2A}) could be given, see e.g. \cite{CAD2009}, however such spectral series can converge extremely rapidly and the higher particle contributions can be almost zero. In the case of free fermions the two-particle form factors give the following,
\begin{equation}\label{IsingExact}
S_A^{(2)}(r)=\frac{1}{8}\log{\xi}+U^{(2)}-\log(1+\mathcal{I}(rM))+\ldots
\end{equation}
with \cite{CCAD2008}
\begin{equation}
\mathcal{I}(r)=\frac{1}{4\pi^2}\int_{-\infty}^\infty\frac{K_0(2r\cosh(x/2))}{\cosh(x/2)}dx
\end{equation}
and
\begin{equation}
U^{(2)}=-2\log\langle\mathcal{T}\rangle\approx -0.115.
\end{equation}
We compared this approximation to the exact result \cite{CFH2005} available trough solving an ODE and found that indeed Eq. (\ref{IsingExact}) reproduces the exact result very precisely for $r\gtrsim 0.01\xi$.

\subsection{Massless theories}
In massless relativistic QFTs we can directly calculate the path integral (\ref{PIRiemann}) exploiting conformal invariance and using a consistent regularization scheme. In fact, we can go further and treat excited states on the same footing thanks to the state-operator correspondence, i.e. implementing the excited state by inserting local operators in the far (Euclidean) past and future.

Following \cite{HLW1994,CC2004,ABS2011,P2014} we use a series of conformal maps to obtain the excited state exponentialized R\'enyi entropy in terms of four point functions on the complex plane. First the finite volume physical manifold of a cylinder ($0\leq x\leq L$, $\tau\in\mathbb{R}$) is mapped to the complex plane,
\begin{equation}
\xi=\exp\left\{\frac{2\pi i}{L}(x-i\tau)\right\},\quad
\bar{\xi}=\exp\left\{-\frac{2\pi i}{L}(x+i\tau)\right\}
\end{equation}
so that the insertion points of the operators implementing the excited states will be in 0 and $\infty$. We identify $\text{Tr}_A\rho_A^2$ with 
\begin{equation}\label{fourR}
\text{Tr}_A\rho_{A,\Psi}^2=N\langle\Psi(0_1,0_1)\Psi(0_1,0_1)^\dagger\Psi(0_2,0_2)\Psi(0_2,0_2)^\dagger\rangle_\mathcal{R}
\end{equation}
where the indices refer to coordinates on the two different Riemann sheets. In CFT the normalization can readily be determined as the ratio $Z_n/Z_1^n$ of partition functions of the $n$-sheeted and 1-sheeted Riemann surfaces irrespective of the particular state. In finite volume this leads to \cite{HLW1994,CC2004}
\begin{equation}
N=\left[\frac{L}{\pi\varepsilon}\sin(\pi d)\right]^{-c/4}
\end{equation}
with a UV regularization parameter $\varepsilon$ and we introduced $d=r/L$ the relative subsystem size.

Now using the unifying map
\begin{equation}
z=f_{d}(\xi)=\sqrt{\frac{e^{-i\pi d}\xi-1}{\xi-e^{-i\pi d}}}
\end{equation}
prescribed to take the first sheet to the principal branch and the second to the other branch of the square root, we express the four point function (\ref{fourR}) by a four point function on the plane,
\begin{equation}\label{fourAd}
\text{Tr}_A\rho_{A,\Psi}^2=N\langle\prod_{i=1}^2(\mathcal{T}_{f_d}\Psi(0_i,0_i))(\mathcal{T}_{f_d}\Psi(0_i,0_i)^\dagger)\rangle_\mathbb{C}
\end{equation}
We have to track the transformation of the inserted operators under this mapping. We introduce the notation
\begin{equation}
\mathcal{T}_{f}\Psi(\xi,\bar{\xi})=U_f\Psi(\xi,\bar{\xi})U_f^{-1}
\end{equation}
for the transformed operators. The generic transformation rule of descendant operators was worked out in Ref. \cite{G1994} and we recall it in Appendix A. Using the composition rule of conformal transformation we can eliminate the adjoints from (\ref{fourAd}) and finally observing the equality $f_d(1/\xi)=f_{(-d)}(\xi)$ we arrive at \cite{P2014}
\begin{equation}\label{four}
\text{Tr}_A\rho_{A,\Psi}^2=N\langle\prod_{i=1}^2(\mathcal{T}_{f_d}\Psi(0_i,0_i))(\mathcal{T}_{f_{-d}}\Psi(0_i,0_i))\rangle_\mathbb{C}
\end{equation}
where we supposed that the adjoint field is the same as the original one.

This four point function can now be evaluated by a systematic approach that generates all descendant four point function by exploiting the conformal Ward identities 
and determining the differential operator producing it from four point functions of primary fields. For chiral fields this leads to
\begin{equation}
N\mathcal{D}_{d}(z_1,z_2,z_3,z_4)\langle\Phi(z_1)\Phi(z_2)\Phi(z_3)\Phi(z_4)\rangle
\end{equation}
with $z_{1,2}=-z_{3,4}=\exp(\pm i\pi d/2)$.
The generalization for fields with both chirality is straightforward.

\section{Truncated conformal space approach}

We will take advantage of the success of the general framework for R\'enyi entropies of excited states in conformal field theories and we will approximate the states of the massive theory with a linear combination of excited states in the underlying CFT through the TCSA. 

To outline the approach let us again consider periodic boundary conditions. On the corresponding Euclidean space-time cylinder of circumference $L$ the Hamiltonian takes the form,
\begin{equation}
H=H_{\text{CFT}}+\lambda\int_0^L dx\,\Phi(0,x)
\end{equation}
where $H_{\text{CFT}}$ is the Hamiltonian of the UV limiting CFT. We will diagonalize this Hamiltonian on the conformal Hilbert space $\mathcal{H}=\{\vert\Psi_i\rangle\}$. This can be written as
\begin{equation}
\mathcal{H}=\bigoplus_i \mathcal{V}_{\Delta_i}\otimes\mathcal{V}_{\bar\Delta_i}
\end{equation}
where $\mathcal{V}_{\Delta}$ is the representation of the Virasoro algebra with highest weight $\Delta$. Supposing a discrete set of primary operators we get a finite matrix problem when a truncation is imposed on the conformal space by throwing away less important states, e.g. with energy greater than some $E_{\text{cut}}$. We can write the TCSA Hamiltonian as \cite{tcsa,PT2008}
\begin{equation}
H^{\text{TCS}}_{ij}=\frac{2\pi}{L}\left[\left(\Delta_i+\bar\Delta_i-\frac{c}{12}\right)\delta_{ij}+\frac{\lambda L^{2-2\Delta}}{(2\pi)^{1-2\Delta}}(G^{-1}B)_{ij}\right]
\end{equation}
where $G=\langle\Psi_i\vert\Psi_j\rangle$ and $B_{ij}=\langle\Psi_i\vert\Phi(1,1)\vert\Psi_j\rangle_\mathbb{C}$.
In order to diagonalize this matrix we need the three-point functions, $B_{ij}$, which are exactly calculable in a systematic way using the conformal Ward identities (see e.g. \cite{K2005} where this calculation is detailed with a superconformal extension). The diagonalization yields the following
\begin{equation}
H^{\text{TCS}}\vert\Psi^{(r)}\rangle=E^{(r)}\vert\Psi^{(r)}\rangle,\quad\vert\Psi^{(r)}\rangle=\sum_i c_i\vert\Psi_i\rangle
\end{equation}

Application to R\'enyi entropies is facilitated by the state-operator identification
\begin{equation}
\vert\Psi^{(r)}\rangle=\sum_i c_i\vert\Psi_i\rangle\quad\leftrightarrow\quad
\sum_i c_i\Psi_i(0,0)
\end{equation}
We simply insert this operator into (\ref{four}) and perform the sum
\begin{multline}
\text{Tr}_A\rho_{A,\Psi}^2=N\sum_{i_1j_1i_2j_2}c_{i_1}c_{j_1}c_{i_2}c_{j_2}\times\\ \times\langle\prod_{k=1}^2(\mathcal{T}_{f_d}\Psi_{i_k}(0_{k},0_{k}))(\mathcal{T}_{f_{-d}}\Psi_{j_k}(0_{k},0_{k}))\rangle_\mathbb{C}
\end{multline}
It is clear that the transformation of all the operators in the operator space and the four point functions of all combinations of them will be required to find the R\'enyi entropy. The transformation rules can be worked out in a straightforward way (see e.g. Appendix \ref{appA}) but obtaining the four-point functions of all (inequivalent) combinations of local fields seems at first extremely challenging. Our experience indeed suggested that an exact approach to the four point functions is necessary to explore the entire $0\leq d\leq 1$ domain, and achieving any progress is possible only when a systematic computer algebraic implementation of these calculations is used.

\section{Deformations of massless free fermions}

We will study deformations of the $c=1/2$ Ising CFT. First let us recall the Hilbert space structure. It consists of the following Verma modules,
\begin{equation}
\mathcal{V}_{0},\quad\mathcal{V}_{1/16},\quad\mathcal{V}_{1/2}
\end{equation}
and the Hilbert space is
\begin{equation}
\mathcal{H}=\mathcal{V}_{0}\otimes\mathcal{V}_{0}+\mathcal{V}_{1/16}\otimes\mathcal{V}_{1/16}+\mathcal{V}_{1/2}\otimes\mathcal{V}_{1/2}
\end{equation}
The corresponding spin zero fields are the identity $1$, the magnetization $\sigma$ and the energy density $\varepsilon$ operators.

To calculate the R\'enyi entropies after determining numerically the representations of massive states on the CFT basis, we also need all the $n$-point functions up to $n\leq4$ between all possible operators. The primary operators have the following $n$-point functions. One point functions are vanishing except for that of the identity,
\begin{equation}
\langle\phi_i(z,\bar z)\rangle=\delta_{i0}
\end{equation}
Two point functions are fixed by normalization,
\begin{equation}
\langle\phi_i(z_1,\bar z_1)\phi_j(z_2,\bar z_2)\rangle=\frac{\delta_{ij}}{(z_1-z_2)^{2\Delta_i}(\bar z_1-\bar z_2)^{2\bar \Delta_i}}
\end{equation}
Three-point functions are fixed by global conformal maps and the structure constants,
\begin{align}
&\langle\phi_i(z_1,\bar z_1)\phi_j(z_2,\bar z_2)\phi_k(z_3,\bar z_3)\rangle\nonumber\\
&\qquad=C_{ijk}(z_1-z_2)^{\Delta_k-\Delta_i-\Delta_j}(\bar z_1-\bar z_2)^{\bar \Delta_k-\bar \Delta_i-\bar \Delta_j}\nonumber\\
&\qquad\qquad\times(z_1-z_3)^{\Delta_j-\Delta_i-\Delta_k}(\bar z_1-\bar z_3)^{\bar \Delta_j-\bar \Delta_i-\bar \Delta_k}\nonumber\\
&\qquad\qquad\times(z_2-z_3)^{\Delta_i-\Delta_j-\Delta_k}(\bar z_2-\bar z_3)^{\bar \Delta_i-\bar \Delta_j-\bar \Delta_k}
\end{align}
where the only nonzero nontrivial structure constants are $C_{\{\varepsilon\sigma\sigma\}}=1/2$ (with arbitrary permutations of the three operators). The four point functions are the following
\begin{align}
&\langle\varepsilon(z_1,\bar z_1)\varepsilon(z_2,\bar z_2)\varepsilon(z_3,\bar z_3)\varepsilon(z_4,\bar z_4)\rangle=\vert\mathcal{F}_1(\{z_i\})\vert^2\\
&\langle\varepsilon(z_1,\bar z_1)\varepsilon(z_2,\bar z_2)\sigma(z_3,\bar z_3)\sigma(z_4,\bar z_4)\rangle=\vert\mathcal{F}_2(\{z_i\})\vert^2\\
&\langle\sigma(z_1,\bar z_1)\sigma(z_2,\bar z_2)\sigma(z_3,\bar z_3)\sigma(z_4,\bar z_4)\rangle\nonumber\\&\qquad\qquad\qquad\quad\qquad=\vert\mathcal{F}_{3}^+(\{z_i\})\vert^2+\vert\mathcal{F}_{3}^-(\{z_i\})\vert^2
\end{align}
with the conformal blocks
\begin{align}
&\mathcal{F}_1(\{z_i\})=\frac{1}{z_{12}z_{34}}+\frac{1}{z_{14}z_{23}}+\frac{1}{z_{24}z_{31}}\\
&\mathcal{F}_2(\{z_i\})=\frac{z_{13}z_{24}-z_{12}z_{34}/2}{z_{12}z_{13}^{1/2}z_{14}^{1/2}z_{23}^{1/2}z_{24}^{1/2}z_{34}^{1/8}}\\
&\mathcal{F}_{3}^{\pm}(\{z_i\})=\left(\frac{z_{13}z_{24}}{z_{12}z_{14}z_{23}z_{34}}\right)^{1/8}\sqrt{1\pm\sqrt{\frac{z_{12}z_{34}}{z_{13}z_{24}}}}
\end{align}
$n$-point functions of descendant can be calculated from the primary $n$-point functions using conformal Ward identities. This calculation is based on the same principles as that of the descendant three-point functions already needed for the TCSA Hamiltonian. However  in case of four-point functions the complexity increases far quicker and we need to retain the functional dependence on the anharmonic ratio throughout the algebraic manipulations. In fact, using the Virasoro algebra we could go up only until descendance level 5 using the available computational resources.

\begin{figure}[t]
\centering
\includegraphics[width=0.95\columnwidth]{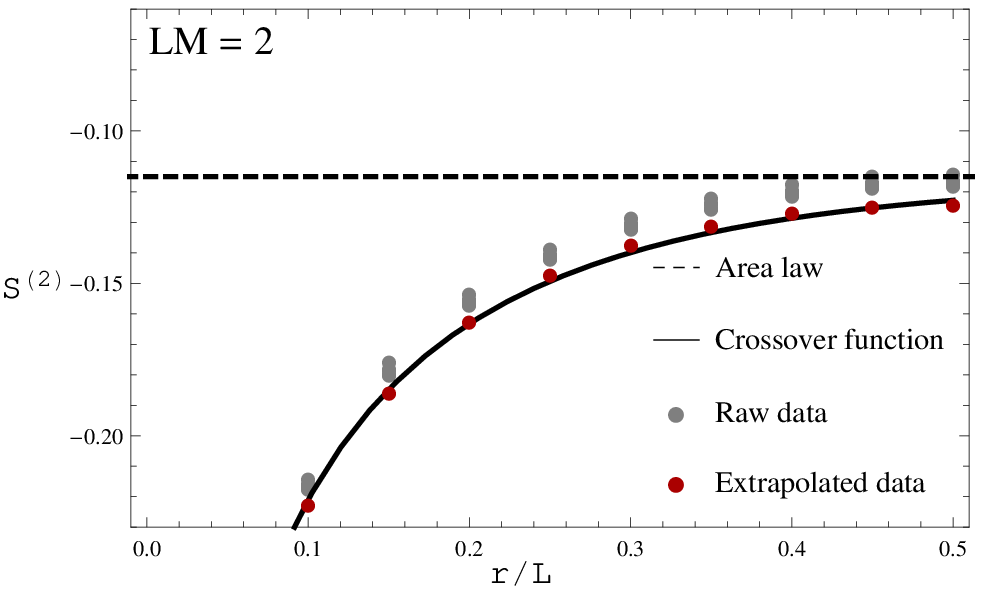}\\[0.3cm]

\includegraphics[width=0.95\columnwidth]{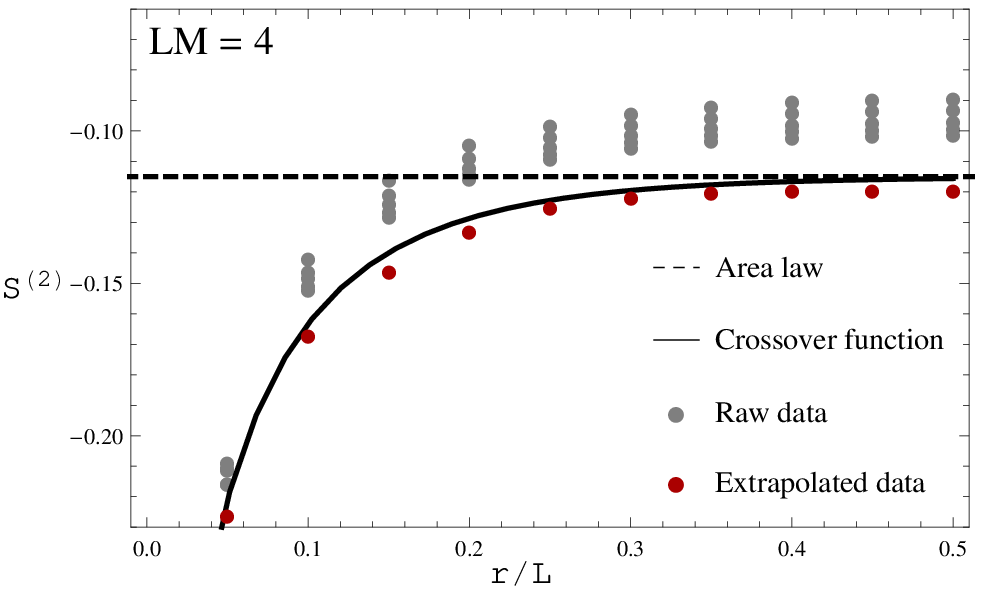}\\[0.3cm]

\includegraphics[width=0.95\columnwidth]{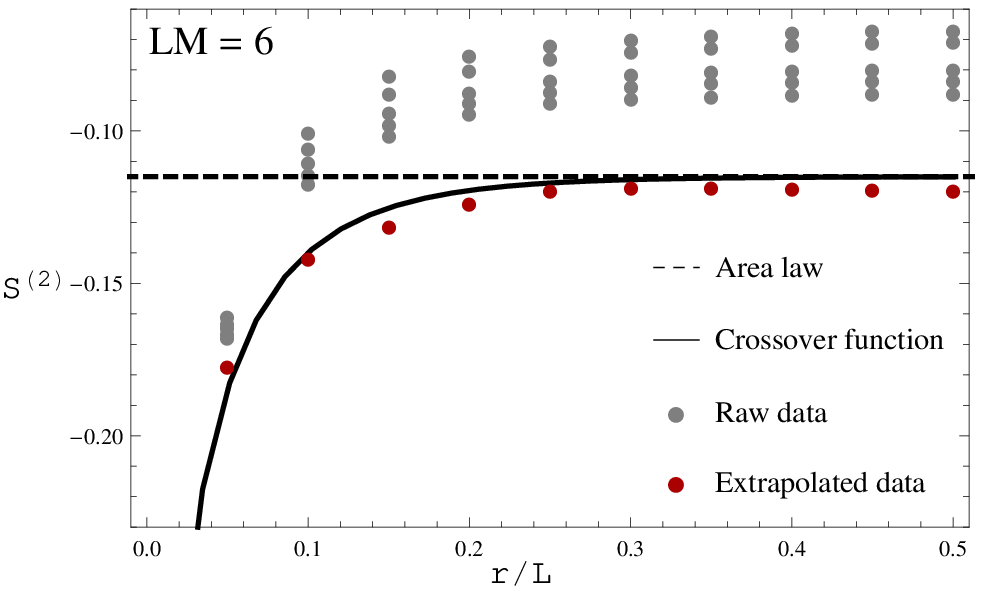}
\caption{Crossover of the 2nd R\'enyi entropy from logarithmic to area law as a function of relative subsystem size $r/L$ at different finite system sizes $LM=2$, 4, 6, and comparison to the prediction in the infinite system.}\label{IsingGS}
\end{figure}

\subsection{Ground state in the massive deformation}

The massive deformation, i.e. giving a finite mass to the fermions, corresponds to the perturbation by the operator $\varepsilon$,
\begin{equation}
H  =  H_{\text{CFT}} + t\int_{0}^{L}\varepsilon(0,x)dx
\end{equation}
Since this model is that of free fermions it is possible to connect the physical mass scale $M$ to the bare coupling by an exact relation $t/M=0.159155$ \cite{f1994}.

We made numerical TCSA calculations to obtain the ground state of the massive deformation of the Ising model. We used a mere 30 state basis corresponding to an energy cutoff $E_\text{cut}=10\,M$. Calculations are further simplified, since the Hamiltonian is block diagonal on the subspaces defined by descendants of $\vert\sigma\rangle$ and those of $\vert0\rangle$ and $\vert\varepsilon\rangle$ (in the lattice picture this is because fermion number is conserved irrespective of $t$).

Our results for three settings of the volume $L$ are exhibited in Fig. \ref{IsingGS}. We show results obtained in the subspace based on $\vert0\rangle$ and $\vert\varepsilon\rangle$. Surprisingly, already in these small settings of the volume we could recover the infinite volume exact crossover function \cite{CFH2005} from logarithmic to massive behavior. We can surmise that the remarkable agreement already at $LM=2$ is due to exponential suppression of finite volume effects. When increasing the volume we see a systematic departure from the prediction which can be explained as a cutoff effect, indeed it is mitigated by an RG improvement according to $S^{(2)}_{E_{\text{cut}}}-S^{(2)}_{\infty}\sim E_{\text{cut}}^{-\alpha}$, where $\alpha=1$ was used. (Similar RG improvements for form factors and vacuum expectation values were discussed in e.g. \cite{2Dtcsa,STW2013}.) From these results we conclude that the TCS approach to calculate the R\'enyi entropy is feasible and reliable.

\subsection{Excited states and TFIM}
To go further, we can check the TCSA against results coming from a spin chain realization of the massive free fermions, the Transverse Field Ising Model. This lattice model has the Hamiltonian
\begin{equation}
H=-\frac{1}{2}\sum_{j=1}^N\sigma_j^x\sigma_{j+1}^x+h\sigma_j^z
\end{equation}
and it can be diagonalized in terms of noninteracting fermions. We will especially be interested in the performance in reproducing the R\'enyi entropies of excited states. Our approach treats the excited states on the same footing as the ground state and a priori we expect comparable performance for both cases. The spin chain data for the R\'enyi entropies come from using the approach discussed in Refs. \cite{Latorre, Berganza}.

\begin{figure}[t]
\centering
\includegraphics[width=0.9\columnwidth]{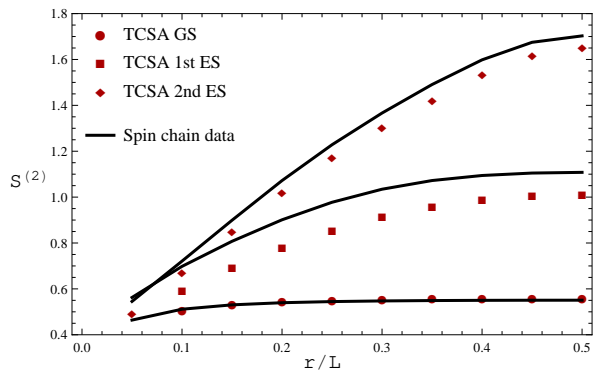}
\caption{2nd R\'enyi entropies of the ground state and first two excited states of the TFIM close to the critical point with $h=0.98$ and $N=200$ and predictions calculated from the TCS ($LM=4$). The TCSA data was obtained at an energy cutoff $E_{\text{cut}}=10\,M$. The nonuniversal constant was identified using the ground state to be $c'_1=0.66$.}\label{figexc}
\end{figure}

The relationship between the parameters of the lattice and continuum versions is given by the formula $LM=N(1-h)$. Then scaling with $N\to\infty$ and $(1-h)\to0$ we can approach the regime of massive QFT where the TCSA can provide data.

To compare the spin chain results to field theory we have to remember that there is a nonuniversal constant correction contribution to the entropies coming from introducing the lattice spacing, 
\begin{equation}
(S^{(2)}_A)_\text{lattice}=(S^{(2)}_A)_\text{QFT}+c'_1
\end{equation}

The results for the first three states are shown in Fig. \ref{figexc}. Here we see more serious departure from the predictions and again we understand this as a cutoff effect. It is interesting that both with the ground state and excited states the correction due to cutoff seems a constant function in the subsystem size.

\subsection{TFIM in a longitudinal field}
When the above spin chain is placed in a longitudinal field a plethora of interesting physics can be observed ranging from the realization of Zamolodchikov's famous $E_8$ model \cite{Zamolodchikov1991} to nonintegrable models with particle confinement.

We studied the integrable point \cite{Zamolodchikov1991}, which in the scaling limit corresponds to a different deformation of the massless Ising model,
\begin{equation}
H  =  H_{\text{CFT}} + s\int_{0}^{L}\sigma(0,x)dx
\end{equation}
where the mass scale is known to be $s=0.06203236\,M^{15/8}$ \cite{f1994}. Compared to the previous deformation this has the advantage being realized by an extremely relevant operator. From this we expect better convergence rates a priori. However, the downside is that while in the absence of a longitudinal field the states could be built using only the tower of $\sigma$ or equivalently, the towers $1$ and $\varepsilon$, here all CFT states contribute to the expansions and we need \emph{all} four point functions to assemble the R\'enyi entropies. Because of this complication and numerical instabilities  we could only achieve a cutoff of $E_\text{cut}=8M$, or equivalently 18 states.

\begin{figure}[!h]
\centering
\includegraphics[width=0.95\columnwidth]{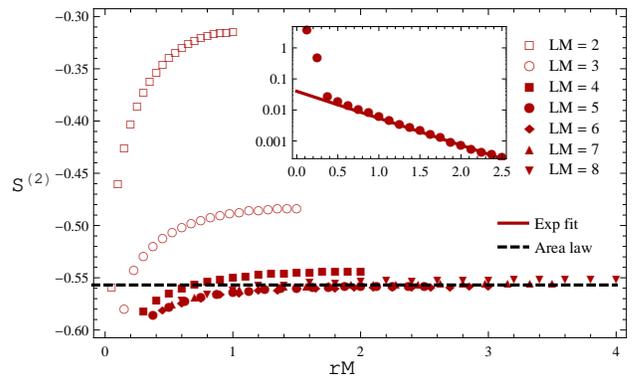}
\caption{2nd R\'enyi entropy of the ground state of the $E_8$ model. The TCSA data was obtained at an energy cutoff $E_{\text{cut}}=8\,M$ using 18 basis states. The various TCSA datasets correspond to different finite volumes $LM=2$, 3, 4, 5, 6, 7, 8 and the graphs are shown only up to half subsystem size. The inset depicts the difference between the saturation value and TCSA data at $LM=5$ versus an exponential fit $e^{-2rM}$ as predicted by the form factor approach (\ref{S2A}).}\label{figE8}
\end{figure}

Results for the R\'enyi entropies of the ground state are shown in Fig. \ref{figE8}. First of all, we observe a pronounced dependence on the volume, however we also observe that at $LM=5$ the finite volume effects are gone and the entropy function becomes independent of $L$. (Slight truncation effects can be seen in the form of a drift with increasing volumes.) We identify the saturation value $S^{(2)}\to -0.557$. In the approach of this value we recognize the exponential decay $S^{(2)}(r)-S^{(2)}(\infty)\sim e^{-2rM}$ predicted by (\ref{S2A}) and confirming that the one-particle form factor of the twist field vanishes. This is in contrast with the fact that other operators, e.g. the trace of the stress energy tensor, have non-vanishing one-particle form factors \cite{Delfino1995}.

\section{Conclusions}

We explored the applicability of the truncated conformal space approach to extract entanglement entropies in energy eigenstates of massive relativistic quantum field theories. We focused on the second R\'enyi entropy and argued that in possession of all four point functions in a truncated conformal space, it is possible to determine them numerically in models realized as relevant perturbations of a conformal field theory. 

We applied our approach to deformations of the $c=1/2$ massless Majorana fermions corresponding to the scaling limit of the transverse field Ising model in a longitudinal field. We presented results for the two integrable points. In the massive free fermion point we could recover the exact crossover function \cite{CFH2005} and for the first two excited states obtained encouraging agreement with data coming from a lattice realization. Finally, we calculated the R\'enyi entropy for the ground state in the other integrable point. We could numerically determine the saturation value and determined from its approach that, interestingly, the twist fields in this model have vanishing one-particle form factors.

While we could obtain reliable and relatively precise results, we were forced to use extremely low cutoffs when truncating the conformal space. The limiting factor is that in our approach four-point function of all combinations of all the local operators (up to the truncation) are needed. It would be very interesting to develop a calculation method to obtain such four point functions in a more efficient way, which could be done e.g. in the $c=1$ model using the free boson representation. Another exciting question is the applicability to other measures of entanglement, e.g. the von Neumann entropy and for mixed states negativity, or to study nonequilibrium evolution of entanglement entropy.

\begin{acknowledgments}
I am grateful to G. Takacs, G. Mussardo, G. Sierra, P. Calabrese, B. Doyon and O. Castro-Alverado for useful discussions. Financial support of the Hungarian Academy of Sciences through grant No. LP2012-50 and a postdoctoral fellowship is acknowledged.
\end{acknowledgments}

\appendix

\section{Conformal transformation of generic fields}\label{appA}

The transformation rules for descendant fields is known \cite{G1994} but it is very rarely used and therefore we see it useful to collect the formulas that we used in this work. We will focus on only the chiral part of the fields and will follow the notations of \cite{DiFrancesco}.

Let us consider a generic descendant operator $\mathcal{O}(z)$ inserted in $z$ and corresponding to the state $\mathcal{O}(0)\vert0\rangle$. A generic form for the transformation of the descendant field can be taken as
\begin{equation}\label{A1}
U_{f}\mathcal{O}(z)U_{f}^{-1}=\sum_{(p)}H_{(p)}[f,z)\left[L_{p_{1}}\ldots L_{p_{k}}\mathcal{O}\right](f(z))
\end{equation}
where $L_{j}$ is the $j$-th Virasoro generator (relative to the origin) satisfying the Virasoro algebra
\begin{equation}\label{Viralg}
[L_n,L_m]=(n-m)L_{n+m}+\frac{c}{12}(n^3-n)\delta_{n+m,0}
\end{equation}
In the formula (\ref{A1}) the coefficients can be given using the state-operator correspondence \cite{G1994} and they are implicitly determined by
\begin{align}
\prod_{n=0}^{\infty}e^{R_{n}[f,z)L_{n}}\mathcal{O}(0)\vert0\rangle
=\sum_{(p)}H_{(p)}[f,z)L_{p_{1}}\ldots L_{p_{k}}\mathcal{O}(0)\vert0\rangle
\end{align}
where the coefficients $R_{n}$ are defined recursively as 
\begin{eqnarray}
R_{0}(z) & = & \log f'(z),\\
R_{n}(z) & = & \frac{1}{n+1}(R'_{n-1}(z)-A_{n}(z)),\quad n\geq1\nonumber 
\end{eqnarray}
with the first few $A_{n}(z)$ being 
\begin{align}
&A_{1}=0,\quad A_{2}=R_{1}^{2},\quad A_{3}=0,\quad A_{4}=\frac{3}{2}R_{2}^{2},\\& A_{5}=A_{6}=A_{7}=0
\end{align}

It is easy to check that $R_{2}$ is of course the Schwarzian of $f$ multiplied by $1/3!$, reproducing the familiar transformation law of the stress-energy tensor,
\begin{equation}
U_{f}T(z)U_{f}^{-1}=\left[f'(z)\right]^{2}T(f(z))+\frac{c}{12}\{f,z\}
\end{equation}
Another example is the transformation law of the field $L_{-1}\phi\equiv\partial\phi$:
\begin{equation}\label{eq:L-1}
U_{f}\partial\phi(z)U_{f}^{-1}=\left[f'(z)\right]^{h+1}\partial\phi(f(z))+\frac{hf''(z)}{\left[f'(z)\right]^{1-h}}\phi(f(z)),
\end{equation}
where $\phi$ is a primary field with scaling weight $h$. This can be deduced, independently, also from the transformation rule of primary fields and the chain rule of differentiation.

\end{document}